\documentclass[twoside]{dis04}
\def\runauthor{Stefano Frixione}
\def\shorttitle{NLO QCD and Monte Carlos}
\def\beq{\begin{equation}}
\def\eeq{\end{equation}}
\def\beqn{\begin{eqnarray}}
\def\eeqn{\end{eqnarray}}
\newcommand{\Matrix}{\mbox{${\cal M}$}}
\newcommand\sss{\scriptscriptstyle}

\newcommand\as{\alpha_{\sss S}}         
\newcommand\pt{p_{\sss T}}         
\newcommand\kt{k_{\sss T}}         
\newcommand\nE{n_{\sss E}}
\newcommand\dsdO{\frac{d\sigma}{dO}}
\newcommand\mat{{\cal M}}
\newcommand\GenMCatNLO{{\cal F}_{\mbox{\tiny MC@NLO}}}
\newcommand\GenMC{{\cal F}_{\mbox{\tiny MC}}}

\begin{document}

\title{The inclusion of higher-order QCD corrections\\
into Parton Shower Monte Carlos}

\author{Stefano Frixione}

\address{INFN, Sezione di Genova\\ 
Via Dodecaneso 33, 16146 Genova, Italy\\
E-mail: Stefano.Frixione@cern.ch}

\maketitle

\abstracts{Substantial progress has been made recently in the
formalisms which form the basis of event generators. I discuss
some of the ideas involved, emphasizing the role of higher-order
QCD corrections and their interplay with parton showers
}

\section{Introduction} 
Event Generators (denoted as EvG's henceforth) have been the workhorses
of all modern experiments in high-energy physics. For good reasons:
in spite of being conceptually simple, they provide fairly
good descriptions of the real events occurring in detectors, allowing
experimenters to perform a variety of tasks, from computing efficiencies
to design strategies for achieving given measurements or searches.
On the other hand, EvG's may not be the ideal tools for predicting
the physical observables with high accuracy, something that is needed in 
order to -- say -- extracting the non-computable parameters of the theory 
from data; traditionally, this task is performed by a class of codes that
can be called {\em cross section integrators} (CSI's). In a loose sense, CSI's
can also output events; however, such events can be used only to predict a
limited number of observables (for example, the transverse momentum of
single-inclusive jets) and are not a faithful description of actual events
taking place in real detectors.

Although complementary in nature, EvG's and CSI's are based on the same 
simple description of an elementary process (the {\em hard subprocess}),
which doesn't even need to be a physically-observable one. To clarify
this point, let us consider a gedanken experiment
which, at an imaginary accelerator that collides 45~GeV $u$-quarks 
with 45~GeV $\bar{u}$-quarks, observes a $d\bar{d}$ quark pair produced
through the decay of a $Z^0$. The process of interest is therefore
$u\bar{u} \rightarrow Z^0 \rightarrow d\bar{d}$ at 90~GeV. Any 
theoretical model describing this process must start from the
knowledge of its cross section
\begin{equation}\label{e_xsec_uuZ}
d\sigma(u\bar{u}\rightarrow Z^0 \rightarrow d\bar{d}) ~=~ 
\frac{1}{2\hat{s}}\;
|\Matrix(u\bar{u}\rightarrow Z^0 \rightarrow d\bar{d})|^2 \;
d\Phi_2\,,
\end{equation}
where $d\Phi_2$ is $d\bar{d}$ phase space, $\Matrix$ is the relevant 
matrix element and $\hat{s}$ is the centre-of-mass energy squared.
Equation~(\ref{e_xsec_uuZ}) can be used to write an EvG or a CSI.
After sampling the phase space, i.e. choosing a point in $d\Phi_2$,
one has a complete description of the $u\bar{u}\rightarrow d\bar{d}$
kinematics -- a {\em candidate event}. The candidate 
event's differential cross section (or {\it event weight}) $d\sigma$ is
calculated from eq.~(\ref{e_xsec_uuZ}) and is directly related to the
probability of this event occurring. The information on such a probability
can be exploited in two ways to get the distributions of the physical
observables: (A) the event weights may be used to create
histograms representing physical distributions, or (B) the events may
be {\it unweighted} such that they are distributed according to the
theoretical prediction. Procedure (A) is very simple and is what is
done for CSI's. A histogram of some relevant 
distribution (e.g.\ the transverse momentum of the $d$ quark) is
filled with the event weights from a large number of candidate
events. The individual candidate events do {\em not} correspond to
anything observable but, in the limit of an infinite number of candidate 
events, the distribution is exactly the one predicted by 
eq.~(\ref{e_xsec_uuZ}).  Procedure (B) is a bit more involved, has added
advantages, and is what is done in EvG's. It produces events with
the frequency predicted by the theory being modelled, and the individual
events represent what might be observed in a trial experiment---in this 
sense unweighted events provide a genuine simulation of an experiment.
Strictly speaking, it would be desirable to talk about events only
in the case of unweighted events; it is important to keep in mind that
CSI's, no matter what their specific nature is, cannot output
unweighted events.

What done so far is theoretically well defined, but scarcely useful,
the process in eq.~(\ref{e_xsec_uuZ}) being non physical. In fact:
{\em a)} The kinematics of the process is trivial; the $Z^0$
has transverse momentum equal to zero. {\em b)} Quark beams 
cannot be prepared and isolated quarks cannot be detected.
Items {\em a)} and {\em b)} have a common origin.  In eq.~(\ref{e_xsec_uuZ})
the number of both initial- and final-state particles is fixed, i.e. there
is no description of the radiation of any extra particles. This radiation
is expected to play a major role, especially in QCD, given the strength
of the coupling constant. 

In the case of item {\em a)}, the extra radiation taking place on top 
of the hard subprocess corresponds to considering higher-order corrections 
in perturbation theory. In the case of item {\em b)}, it can be viewed as
an effective way of describing the dressing of a bare quark which 
ultimately leads to the formation of the bound states we observe
in Nature ({\em hadronization}). Thus, any EvG or CSI which aims at 
giving a realistic description of collision processes must include: 
{\em i)} A way to compute exactly or to estimate the effects of
higher-order corrections in perturbation theory. {\em ii)} A way 
to describe hadronization effects.
Different strategies have been devised to solve these problems. They
can be quickly summarized as follows. For higher orders: 
{\bf HO.1)} Compute exactly the result of a given (and usually small) 
number of emissions. {\bf HO.2)} Estimate the dominant effects due to 
emissions at all orders in perturbation theory. For hadronization:
{\bf HAD.1)} Use the QCD-improved version of Feynman's parton model ideas 
(the {\em factorization theorem}) to describe the parton $\leftrightarrow$
hadron transition. {\bf HAD.2)} Use phenomenological models to describe 
the parton $\leftrightarrow$ hadron transition at mass scales where 
perturbation techniques are not applicable.

The simplest way to implement strategy {\bf HO.1)} is to consider only those
diagrams corresponding to the emission of real particles.  Basically, the
number of emissions coincides with the perturbative order in $\as$. This
choice forms the core of {\em Tree Level Matrix Element} generators.  These
codes can be used either within a CSI or within an EvG. A more involved
procedure aims at computing all diagrams contributing to a given perturbative
order in $\as$, which implies the necessity of considering virtual emissions
as well as real emissions. Such {\em N$^k$LO computations} are technically
quite challenging and satisfactory general solutions are known only for the
case of one emission (i.e., NLO). Until recently, these computations
have been used only in the context of CSI's; their use within EvG's is a brand
new field, and I'll deal with it in what follows.

Strategy {\bf HO.2)} is based on the observation that the dominant effects in
certain regions of the phase space have almost trivial dynamics, such that
extra emissions can be recursively described. There are two vastly different
classes of approaches in this context. The first one, called {\em resummation},
is based on a procedure which generally works for one observable at a time
and, so far, has only been implemented in cross section integrators.  The
second procedure forms the basis of the {\em Parton Shower} technique and is,
by construction, the core of EvG's. This procedure is not observable-specific,
making it more flexible than the first approach, but it cannot reach the same
level of accuracy as the first, at least formally.

At variance with the solutions given in {\bf HO.1} and {\bf HO.2}, 
solutions to the problem posed by hadronization always involve
some knowledge of quantities which cannot be computed
from first principles (pending the lattice solution of the theory)
and must be extracted from data. The factorization theorems mentioned
in {\bf HAD.1} are the theoretical framework in which CSI's are
defined. Parton shower techniques, on the other hand, are used
to implement strategy {\bf HAD.2} in the context of EvG's.

\section{Event Generators at TeV Colliders}
As discussed in the previous section, EvG's and those CSI's which are
based upon strategy {\bf HO.2} for the description of higher-order
corrections (i.e. those that implement some kind of resummation)
give exactly the same description for the observables for which the
analytical computations required by the CSI's are feasible\footnote{An
alternative approach to resummation, based on numerical methods, has been 
recently proposed in ref.~\cite{Banfi:2004yd}.}, provided that the logarithmic 
accuracy of the shower and of the resummation is the same. This is basically 
never the case; analytical resummations are more accurate than 
parton showers. In practice, some of the (formally uncontrolled)
higher logarithms sneak in the showers, and the effective resummation
performed by EvG's is seen to give, in many cases, results which are
very close to those obtained with analytical resummation techniques.
For this reason, the so-far unknown solution of the interesting and 
fairly challenging problem of improving the logarithmic accuracy of 
the showers would presumably give only marginal effects in phenomenological
predictions. On the other hand, the improvement in the treatment of
soft emissions at large angles would have a more visible effect, although
on a more restricted class of observables.

The multiple emissions of quarks and gluons performed by the showers change 
the kinematics of the hard subprocess. The $Z^0$ of eq.~(\ref{e_xsec_uuZ})
acquires a non-zero transverse momentum $\pt$ by recoiling against the
emitted partons. Since the parton shower is based upon a collinear
approximation, one must expect the predictions of an EvG for, say,
$\pt(Z^0)>100$~GeV to be completely unreliable. Fortunately, the bulk of
the cross section occurs at much smaller values of $\pt$, where EvG's
do provide a sensible description of the production process. In the
energy range involved in the collider physics program up to now, this 
was sufficient for the vast majority of the experimenters' needs.

The situation has now changed considerably. Tevatron Run II and especially 
LHC will feature very high-energy, high-luminosity collisions, and the
events will have many more energetic well-separated particles/jets than 
before. An accurate description of these is necessary, especially in view
of the fact that signals for many beyond-the-SM models involve
in fact a large number of jets, resulting from the decay chains
of particles of very high mass. The complexity of the LHC 
environment will be such that an incorrect description of the
hard processes {\em may} even jeopardize the discovery potential 
of the machine, and will certainly prevent the experiments from
performing detailed studies of the collision processes.

The collinear nature of the parton shower implies that 
EvG's cannot do well in predicting high-$\pt$ processes.
The fact that the description of the hard process is achieved 
using a leading-order picture, as outlined in the previous
sections, has also a second implication: estimates of the rates
(i.e., of the number of particles to be detected by the experiments)
will be largely underestimated, since many processes have large
K factors. Troubles arise when not only the K factors are large,
but differ sizably between the various processes, since this 
complicates enormously the task of normalizing the signal using the 
background. It should be clear that the K factors needed here are
those relevant to the visible regions of the detectors. It is usually
assumed that the ratios of these is equal to ratios of the fully
inclusive K factors. This crude approximation usually works decently,
but may fail dramatically when a complex kinematics is at play.

The bottom line is that the EvG's, which have been one of the fundamental
building blocks of the very successful collider physics program
of the 80's and the 90's, will not perform well with
the new generation of experiments. They will need either to be
improved, or to be replaced.

The emphasis on large-$\pt$ emissions implies that the only candidates
for the replacement of EvG's are the CSI's that implement exactly the
kinematics of the higher-order QCD corrections, thus performing N$^k$LO 
computations (strategy {\bf HO.1}). Unfortunately, it is at present unknown 
how to cancel systematically, and without any reference to a specific
observable, the infrared and collinear singularities beyond NLO.
Besides, the description of the hadronization phenomena in such
computations is very crude, and cannot match the sophistication
of the hadronization model implemented in EvG's. Furthermore, as
already mentioned, N$^k$LO computations cannot output events, which
is what is absolutely needed.

Barring the possibility of replacing EvG's with something else, the
only solution left is to improve them; the improved EvG's will be
able to predict sensibly the large-$\pt$ emissions, without losing
their capability of treating fairly the low-$\pt$ region, performing
resummations there. Clearly, since the large-$\pt$ region is associated
with higher-order diagrams, the improvement of EvG's will be equivalent
to answering the following question: {\em How can we insert higher-order
QCD corrections into EvG's?} As I will soon discuss, there are two
different, largely complementary ways, to solve this problem.

\section{Matrix Element Corrections and CKKW}
Since the large-$\pt$ emissions are due to the 
real emission diagrams, the first strategy (denoted as
Matrix Element Corrections, MEC henceforth) is that of considering
only these diagrams among those contributing to higher-order QCD 
corrections, in this way neglecting all the diagrams with one
or more virtual loops. In doing so, the possibility is given up 
of including the K factor consistently in the computations.

The starting point for including real emission diagrams in EvG's
is that of computing them efficiently, which includes efficient
samplings of very complex final-state phase spaces.
Fortunately, techniques are known to highly automatize such
computations, which are nowadays performed by specialized 
codes (the Tree Level Matrix Element generators), external to 
proper EvG's and interfaced to them in a standardized way for FORTRAN-based
event generators by the Les Houches Accord (LHA) event
record~\cite{Boos:2001cv} (the LHA standard is supported in C++ by the
HepMC~\cite{Dobbs:2001ck} event record). Tree-level matrix element 
generators can be divided into two broad classes, which I will 
briefly review below; the interested reader can find more information
in ref.~\cite{Dobbs:2004qw}.

The codes belonging to the first class 
feature a pre-defined list of partonic processes. Multi-leg amplitudes 
are strongly and irregularly peaked; for this reason the phase-space 
sampling has typically been optimized for the specific process. The
presence of phase space routines implies that these codes are always
able to output partonic events (weighted or unweighted). Popular
packages are AcerMC~\cite{Kersevan:2002dd}, AlpGEN~\cite{Mangano:2002ea},
Gr@ppa~\cite{Tsuno:2002ce}, MadCUP~\cite{Madcup}.

The codes belonging to the second class may be thought of as automated matrix
element generator authors. The user inputs the initial and final state
particles for a process.  Then the program enumerates Feynman diagrams
contributing to that process and writes the code to evaluate the matrix
element.  The programs are able to write matrix elements for {\it any} tree
level SM process. The limiting factor for the complexity of the events is
simply the power of the computer running the program.  Typically Standard
Model particles and couplings, and some common extensions are known to the
programs.
Many of the programs include phase space sampling routines. As such, they
are able to generate not only the matrix elements, but to use those
matrix elements to generate partonic events (some programs also include
acceptance-rejection routines to unweight these events).
Codes belonging to this class are AMEGIC++~\cite{Krauss:2001iv}, 
CompHEP~\cite{Pukhov:1999gg}, Grace~\cite{Fujimoto:2002sj}, 
MadEvent~\cite{Maltoni:2002qb}.

The use of one of the codes listed above allows one to generate
a final-state configuration made of hard quarks, gluons, and
other non-coloured particles such as Higgs or gauge bosons. This
final state is thus not directly comparable to what is observed in
a detector. A drastic simplification is that of assuming that there is a one 
to one correspondence between hard partons and physical jets. 

However, this 
assumption may cause problems when interfacing these codes to EvG's such as 
HERWIG~\cite{Corcella:2000bw} or PYTHIA~\cite{Sjostrand:2000wi}; a step 
which is necessary in order to obtain more sensible descriptions of the 
production processes. In fact, a kinematic configuration with $n$ final-state
partons can be obtained starting from $n-m$ partons generated by the
tree-level matrix element generator, with the extra $m$ partons provided 
by the shower. This implies that, although the latter partons are generally 
softer than or collinear to the former, there is always a non-zero probability
that the {\em same} $n$-jet configuration be generated starting from 
{\em different} $(n-m)$-parton configurations. Since 
tree-level matrix elements do have soft and collinear singularities, a cut 
at the parton level is necessary in order to avoid them\footnote{It is
actually this cut that defines the ``hardness'' of the primary partons.};
I will symbolically refer to this cut as $y_{cut}$ in what follows.
Physical observables should be independent of $y_{cut}$, but they are not;
the typical dependence is of leading-log nature 
(i.e., $\as^k \log^{2k}y_{cut}$).

To clarify this issue with a simple example, let me consider again the hard 
subprocess of eq.~(\ref{e_xsec_uuZ}), $u\bar{u} \rightarrow Z^0$. One of 
the NLO real contributions to this process is $u\bar{u} \rightarrow Z^0g$.
Events from these two processes should never be blindly combined, since a
fraction of the latter events are already included in the former process via
gluon radiation in the parton shower. Combining the two processes without
special procedures amounts to {\em double counting} some portion of phase
space.

The first approaches to the technique of MEC, which allows one to solve the
double counting problem, limited themselves to the case of at most one extra 
hard parton
wrt those present at the Born level~\cite{Bengtsson:1986et,Seymour:1994df}.
These MEC can be implemented either as a strict partition of phase
space between two processes, or as an event reweighting (re-evaluation of
the event probability using the matrix element) using the higher order tree
level matrix element for the related process. In either case the effect is the
same: the event shapes are dominated by the parton shower in the low-$\pt$
region, the shapes are NLO-like in the high-$\pt$ region, and the total cross
section remains leading order (i.e.\ for our example the total cross section
will be the same as that for $u\bar{u} \rightarrow Z^0$). The trouble with
such versions of MEC is that they can be applied only in a very
limited number of cases, which are relatively simple in terms of 
radiation patterns and colour connections. 

The way in which MEC can be achieved in the general case of $\nE$ extra hard 
partons, with $\nE\ge 1$, has been clarified in ref.~\cite{Catani:2001cc}
for the case of $e^+e^-$ collisions (referred to as CKKW after the
names of the authors). The idea is the following:
{\em a)} Integrate all the
$\gamma^*\to 2+\nE$ ME's by imposing $y_{ij}>y_{cut}$ for any
pairs of partons $i,j$, with 
$y_{ij}=2\min(E_i^2,E_j^2)(1-\cos\theta_{ij})/Q^2$ the interparton
distance defined according to the $\kt$-algorithm.  
{\em b)} Choose statistically an $\nE$, using
the rates computed in {\em a)}.  {\em c)} Generate a $(2+\nE)$-parton
configuration using the exact $\gamma^*\to 2+\nE$ ME, and reweight it
with a suitable combination of Sudakov form factors (corresponding to
the probability of no other branchings). {\em d)} Use the configuration 
generated in {\em c)} as initial condition for a {\em vetoed} shower.  
A vetoed shower proceeds as the usual one, except that it forbids all
branchings $i\to jk$ with $y_{jk}>y_{cut}$ without stopping the
scale evolution. Although the selection of an $\nE$ value has
a leading-log dependence on $y_{cut}$, it can be proved that this
dependence is cancelled up to next-to-next-to-leading logs in physical
observables (i.e., $\as^k \log^{2k-2}y_{cut}$), plus terms suppressed
by powers of $y_{cut}$. It is clear that, in order to be 
internally consistent, matrix elements must be available for any value
of $2+\nE$. In practice, $\nE\le 3$ is a good approximation of $\nE<\infty$.

After CKKW proposed their implementation of MEC for $e^+e^-$ collisions, an 
extension to hadronic collisions has been presented, without formal
proof, in ref.~\cite{Krauss:2002up}; an alternative method for 
colour-dipole cascades has been presented in ref.~\cite{Lonnblad:2001iq}.
There is a considerable freedom in the implementation of the CKKW prescription
in the case of hadronic collisions. This freedom is used to tune (some of)
the EvG's parameters in order to reduce as much as possible the $y_{cut}$
dependence, which typically manifests itself in the form of discontinuities in
the derivative of the physical spectra. A discussion on these issues,
with practical examples of the implementation of CKKW in HERWIG and
PYTHIA, can be found in ref.~\cite{Mrenna:2003if}. CKKW has also been
implemented in SHERPA~\cite{Gleisberg:2004re}; an alternative procedure, 
proposed by Mangano, is being implemented in AlpGEN.

I stress that the complete independence of $y_{cut}$ cannot be achieved; this
would be possible only by including all diagrams (i.e., also the virtual ones)
contributing to a given order in $\as$.

\section{Adding virtual corrections: NLOwPS}
The point made at the end of the previous section appears obvious; it is well 
known, and formally established by the BN and KLN theorems, that the infrared 
and collinear singularities of the real matrix elements are cancelled by the
virtual contributions. One may in fact be surprised by the mild $y_{cut}$
dependence left in the practical implementation of CKKW (see for example
ref.~\cite{Mrenna:2003if}); however, we should keep in mind that parton
showers do contain part of the virtual corrections, thanks to the
unitarity constraint which is embedded in the Sudakov form factors.
However, to cancel exactly the $y_{cut}$ dependence there is no alternative
way to that of inserting the exact virtual contributions to the hard
process considered. In doing so, one is also able to include consistently
in the computation the K factor. It is important to realize that this
is {\em the only way} to obtain this result in a theoretically 
consistent way. The procedure of reweighting the EvG's results to 
match those obtained with CSI's for certain observables must be considered 
a crude approximation (since no CSI is able to keep into account
all the complicated final-state correlations that are present when
defining the cuts used in experimental analyses).

The desirable thing to do would be that of adding the virtual
corrections of the same order as all of the real contributions
to CKKW implementations. Unfortunately, this is unfeasible, for
practical and principle reasons. The practical reason is that,
at variance with real corrections, we don't know how to automatize
efficiently the computations of loop diagrams in the Minkowskian
kinematic region. The principle reason is that there's no known
way of achieving the cancellation of infrared and collinear divergences 
in an universal and observable-independent manner beyond NLO.
We have thus to restrict ourselves to the task of including 
NLO corrections in EvG's; I'll denote the EvG improved in this
way as NLO with Parton Showers (NLOwPS).

The fact that only one extra hard emission can be included in NLOwPS's
is the reason why such codes must be presently seen as complementary
to MEC. When one is interested in a small number of extra emissions,
then NLOwPS's must be considered superior to MEC; on the other hand,
for studying processes with many hard legs involved, such as SUSY
signals or backgrounds, MEC implementations should be used. A realistic
goal for the near future is that of incorporating the complete NLO 
corrections to all the processes with different $\nE$'s in CKKW.

Before turning to a technical discussion on NLOwPS's, let me specify in 
more details the meaning of ``NLO'' in the context on an EvG. To do
so, let me consider the case of SM Higgs production at hadron colliders,
which at the lowest order, ${\cal O}(\as^2)$, proceeds through a loop
of top quarks which is the only non-negligible contribution to the $ggH$ 
effective vertex. When the $\pt$ distribution of the Higgs is studied,
we get what follows:
\beq
\frac{d\sigma}{d\pt}=\left(A\as^2+B\as^3\right)\delta(\pt)+C(\pt)\as^3\,,
\eeq
which means
\beqn
\int_{\pt^{min}}^\infty d\pt\frac{d\sigma}{d\pt}&=&{\cal C}_3\as^3,
\phantom{+{\cal D}_3\as^3}\;\;\;\;\;\;{\pt^{min}>0}
\\*
&=&{\cal D}_2\as^2+{\cal D}_3\as^3,\;\;\;{\pt^{min}=0}\,.
\eeqn
In the language of perturbative computations, the result for $\pt^{min}>0$
would be denoted as LO, that for $\pt^{min}=0$ as NLO. This is not appropriate
for EvG's, since such a naming scheme depends on the observable considered,
and EvG's produce events without any prior knowledge of the observable(s)
which will eventually be reconstructed. Thus, in the context of EvG's,
we generally define N$^k$LO accuracy with $k$ the number of extra (real
or virtual) gluons or light quarks wrt those present at the Born level.

Apart from this, there is a certain freedom in defining NLOwPS's.
I follow here the definitions given in ref.~\cite{Frixione:2002ik},
were the NLOwPS MC@NLO was first introduced:
\begin{itemize}
\item Total rates are accurate to NLO.
\item Hard emissions are treated as in NLO computations.
\item Soft/collinear emissions are treated as in MC.
\item NLO results are recovered upon expansion of NLOwPS results in $\as$.
\item The matching between hard- and soft/collinear-emission regions is smooth.
\item The output is a set of events, which are fully exclusive.
\item MC hadronization models are adopted.
\end{itemize}
The fourth condition above defines the absence of double counting in
NLOwPS's. In other words: {\em An NLOwPS is affected by double counting 
if its prediction for any observable, at the first order beyond the
Born approximation in the expansion in the coupling constant,
is not equal to the NLO prediction.}
According to this definition, double counting may correspond to
either an excess or a deficit in the prediction, at any point in
phase space. This includes contributions from real emission and
virtual corrections.

Let me now consider a generic hard production process, whose nature
I don't need to specify, except for the fact that its LO contribution
is due to $2\to 2$ subprocesses, which implies that real corrections 
will be due to $2\to 3$ subprocesses; these conditions are by no means
restrictive, and serve only to simplify the notation. Let $O$ be an 
observable whose value can be computed by knowing the final-state
kinematics emerging from the hard processes. At the NLO, we can 
write the distribution in $O$ as follows:
\beqn
\left(\dsdO\right)_{subt}&=&
\sum_{ab}\int dx_1\,dx_2\,d\phi_3\,f_a(x_1)f_b(x_2)
\nonumber \\
&&\biggl[{\delta(O-O(2\to 3))}\mat_{ab}^{(r)}(x_1,x_2,\phi_3)+
\nonumber \\
&&\,\,{\delta(O-O(2\to 2))}\Big(\mat_{ab}^{(b,v,c)}(x_1,x_2,\phi_2)
-\mat_{ab}^{(c.t.)}(x_1,x_2,\phi_3)\Big)\biggr].
\label{NLOpred}
\eeqn
Here, $\mat_{ab}^{(r)}$ is the contribution of the real matrix elements,
whereas $\mat_{ab}^{(b,v,c,c.t.)}$ are the contributions of the Born, virtual,
collinear reminders and collinear counterterms; $O(2\to n)$, with $n=2,3$,
is the value of the observable $O$ as computed with 2- and 3-body
final states. The form of eq.~(\ref{NLOpred}) is borne out by the universal 
formalism for cancelling the infrared and collinear divergences proposed
in refs.~\cite{Frixione:1995ms,Frixione:1997np}, upon which MC@NLO is based.
Other equivalent forms could be used at this point, without changing
the conclusions.

In order to predict the distribution of $O$ using an EvG, one computes 
the value of $O$ for each event generated by the shower. The most compact
way of describing how an EvG works is through the generating functional,
which is basically the incoherent sum of all possible showers
\beqn
\GenMC=\sum_{ab}\int dx_1\,dx_2\,d\phi_2\,f_a(x_1)f_b(x_2)
\;{\GenMC^{(2\to 2)}}\mat_{ab}^{(b)}(x_1,x_2,\phi_2),
\label{GenMC}
\eeqn
where $\GenMC^{(2\to 2)}$ is the generating functional for parton-parton
scattering, with a $2\to 2$ configuration as a starting condition for the
showers.

In the attempt of merging NLO and EvG, we observe that in eqs.~(\ref{NLOpred})
and~(\ref{GenMC}) the short distance matrix elements serve to determine the
normalization of the results, and the hard process kinematics. Such kinematics
configurations are evolved by the showers $\GenMC^{(2\to 2)}$ in 
eq.~(\ref{GenMC}), and the resulting final states eventually used to
compute the value of $O$. A similar ``evolution'' is performed in the
context of the NLO computations by the $\delta$ functions appearing
in eq.~(\ref{NLOpred}); clearly, the evolution is trivial in this case.
However, this suggests that the incorporation of NLO results into EvG's
may simply amount to replacing in eq.~(\ref{NLOpred})
$\delta(O-O(2\to n))$ with $\GenMC^{(2\to n)}$,
i.e. with the generating functionals of the showers whose initial conditions
are $2\to 2$ and $2\to 3$ hard kinematics configurations. It should be 
stressed that this strategy, that I'll call the {\em naive NLOwPS 
prescription}, actually works at the LO, since eq.~(\ref{GenMC}) can be 
obtained from eq.~(\ref{NLOpred}) following this prescription, if terms 
beyond LO are dropped from the latter equation.

Unfortunately, things are more complicated than this. Basically, when
$\GenMC^{(2\to 2)}$ acts on $\mat_{ab}^{(b)}$ in the analogue of 
eq.~(\ref{NLOpred}) obtained by applying the naive NLOwPS prescription, 
it generates terms that contribute to the NLO 
prediction of $O$, which are not present in eq.~(\ref{NLOpred}). 
According to the definition given above, this amounts to double
counting. Furthermore, the weights associated with $\GenMC^{(2\to 2)}$
and $\GenMC^{(2\to 3)}$ (i.e., the coefficients multiplying 
$\delta(O-O(2\to 2))$ and $\delta(O-O(2\to 3))$ in eq.~(\ref{NLOpred})
respectively) are separately divergent. These divergences are known
to cancel thanks to the KLN theorem and the infrared safeness of $O$;
however, this happens efficiently in the case of the NLO computations,
thanks to the fact that the final-state configurations with which the
values of $O$ are computed coincide with the hard configurations. This
is not the case when the showers are attached, since the evolutions
implicit in $\GenMC^{(2\to 2)}$ and $\GenMC^{(2\to 3)}$ are not correlated
(and must not be so). This means that the naive prescription outlined
above, apart from double counting, requires an infinite amount of CPU
time in order for the cancellation of the infrared divergences to occur.

I'll now show how these problems are solved in the context of
MC@NLO~\cite{Frixione:2002ik,Frixione:2003ei}. We observe that,
if the shower evolution attached to the Born contribution in the
naive prescription results in spurious NLO terms, one may try to
remove ``by hand'' such terms. Denoting by 
$\mat_{{\cal F}(ab)}^{\mbox{\tiny(MC)}}$ the terms that we'll actually
remove, the following equation holds:
\beq
\mat_{{\cal F}(ab)}^{\mbox{\tiny(MC)}}=
\GenMC^{(2\to 2)}\mat_{ab}^{(b)}+{\cal O}(\as^2\as^b),
\label{MCcntdef}
\eeq
where $\as^b$ is the perturbative order corresponding to the Born
contribution. Clearly, eq.~(\ref{MCcntdef}) leaves lot of freedom
in the definition of $\mat_{{\cal F}(ab)}^{\mbox{\tiny(MC)}}$
(which I denote as {\em MC counterterms}), in that all terms
of NNLO and beyond are left unspecified. In MC@NLO, we defined
the MC counterterms using eq.~(\ref{MCcntdef}), and requiring all
terms beyond NLO to be zero. With this, we define the MC@NLO
generating functional as follows:
\beqn
\GenMCatNLO&=&\sum_{ab}\int dx_1\,dx_2\,d\phi_3\,f_a(x_1)f_b(x_2)
\label{GenMCatNLO}
\\
&&\biggl[{\GenMC^{(2\to 3)}}\left(\mat_{ab}^{(r)}(x_1,x_2,\phi_3)
-{\mat_{ab}^{\mbox{\tiny(MC)}}(x_1,x_2,\phi_3)}\right)+
\nonumber \\
&&\,\,{\GenMC^{(2\to 2)}}\Big(\mat_{ab}^{(b,v,c)}(x_1,x_2,\phi_2)
-\mat_{ab}^{(c.t.)}(x_1,x_2,\phi_3)+
{\mat_{ab}^{\mbox{\tiny(MC)}}(x_1,x_2,\phi_3)}\Big)\biggr].
\nonumber 
\eeqn
Eq.~(\ref{GenMCatNLO}) is identical to what one would have got by
applying the naive NLOwPS prescription discussed above to eq.~(\ref{NLOpred}), 
except for the fact that the short-distance coefficients
have been modified by adding and subtracting the MC counterterms;
for this reason, MC@NLO is said to be based upon a modified subtraction
method. At the first glance, it may appear surprising that the MC
counterterms have been added twice, with different signs, since their
role is that of eliminating the spurious terms arising from the
evolution of the Born term. However, this is what they do indeed.
In fact, the evolution of the Born term also includes a contribution
due to the so-called non-branching probability, i.e. the probability
that nothing happens. This corresponds to a would-be deficit of the
naive NLOwPS prediction, which is taken into account by our definition
of double counting.

Remarkably, the solution of the problem of double counting also
solves the problem of the cancellation of the infrared and collinear
divergences in a finite amount of time. In fact, the weights attached
to the two generating functionals on the r.h.s. of eq.~(\ref{GenMCatNLO})
are now separately finite locally in the phase space. This is so since
the showers are constructed to reproduce the behaviour of the collinear
emissions as predicted by perturbation theory, and this in turn implies
that the MC counterterms locally match the singular behaviour of the
real matrix elements, hence the name ``counterterms'' (there are subtleties
due to the peculiar treatment of soft emissions in showers, which are
technically too involved to be discussed here; the interested reader
can find all the details in ref.~\cite{Frixione:2002ik}).
This fact also implies that MC@NLO produces events identical in
nature to those of standard EvG's, since unweighting can be performed
at the level of short-distance contributions. As a consequence, the
convergence properties (i.e., the smoothness of the physical distributions)
are much better than those of the corresponding NLO codes; typically, to
achieve the same level of fluctuations, MC@NLO has to sample the phase
space about 50 times less than the NLO code from which it is derived.
This pattern is followed by all of the processes so far implemented in 
MC@NLO, whose (growing) list can be found with the package at 
{\tt http://www.hep.phy.cam.ac.uk/theory/webber/MCatNLO/}.

An important point to stress is that the computation of the MC counterterms
requires a detailed knowledge of what the EvG does when performing the
shower. This means that the MC counterterms are specific to a given
Monte Carlo implementation: those corresponding to HERWIG differ from
those corresponding to PYTHIA. Presently, MC@NLO can only be interfaced
to HERWIG, since only the MC counterterms relevant to HERWIG have been
computed. It is also worth mentioning that the form of the MC counterterms
doesn't depend on the hard process considered; thus, their computation 
is performed once and for all. A second point is that NLOwPS's are in
general not positive definite, i.e. a fraction of the generated events
will have negative weights. Fortunately, this fraction is fairly small,
and future work may lead to its further reduction.

In spite of attracting a considerable amount of theoretical interest
in the past few years, at the moment there are only a couple of codes,
plus MC@NLO, that can be used to produce actual events in hadronic collisions.
Phase-space veto has been introduced in ref.~\cite{Dobbs:2001dq}, elaborating
on an older idea presented in ref.~\cite{Baer:1991qf}, and applied to
$Z^0$ production. The approach is interesting since no negative-weight
events are produced. However, as shown in ref.~\cite{Frixione:2002ik},
this is obtained at the price of double counting in certain
regions of the phase space. Although the practical impact of such
double counting seems to be modest for the physical process considered,
it remains to be seen how the method can be generalized in order to
treat processes more complicated from the point of view of kinematics
and colour configurations. The code grcNLO~\cite{Kurihara:2002ne} is
characterized by the numerical computation of all the matrix elements
involved. In order not to do double counting, the short distance cross
sections have to be interfaced with an {\em ad-hoc} shower, i.e., the
interfacing with HERWIG or PYTHIA does produce double counting. The method has
so far been applied to $Z^0$ production, and efforts are being made in order
to implement $Z^0+1$~jet production.

\enlargethispage*{10pt}
\section{Conclusions}
In the past few years, many significant advances have been achieved
in the theory and implementation of the event generators. Although 
a considerable amount of work remains to be done, it is fair to say
that the codes of the new generation will be up to the challenge
posed by Tevatron Run II and LHC physics.

\end{document}